\documentclass[aps,prd,twocolumn,preprintnumbers,superscriptaddress,dblfloatfix,nofootinbib]{revtex4-2}
\usepackage[utf8]{inputenc}
\usepackage{lmodern}
\usepackage{amsmath,amssymb,mhchem}
\usepackage{graphicx}
\usepackage{url}
\usepackage{color}
\usepackage{enumitem}
\usepackage{subfigure}
\usepackage[dvipsnames]{xcolor}
\usepackage[colorlinks=true,breaklinks=true]{hyperref}
\hypersetup{allcolors=[rgb]{0.0 0.0 0.6},linkcolor=[rgb]{0.75 0.05 0.05}}
\usepackage{bm}
\usepackage{epsfig}
\usepackage{amsmath}
\usepackage{amssymb}
\usepackage{slashed}
\usepackage{color}
\usepackage{accents}
\usepackage[dvipsnames]{xcolor}
\usepackage[colorlinks=true,breaklinks=true]{hyperref}
\hypersetup{allcolors=[rgb]{0.0 0.0 0.6},linkcolor=[rgb]{0.75 0.05 0.05}}
\usepackage[export]{adjustbox}

\newcommand{\beq}{\begin{equation}}
\newcommand{\eeq}{\end{equation}}
\def\bea{\begin{eqnarray}}
\def\eea{\end{eqnarray}}
\newcommand{\bei}{\begin{itemize}}
\newcommand{\eei}{\end{itemize}}
\newcommand{\bmat}{\begin{matrix}}
\newcommand{\emat}{\end{matrix}}

\def\={\,=\,}
\def\+{\,+\,}
\def\-{\,-\,}

\begin{document}

\title{Primordial Black Hole Reformation in the Early Universe}

\author{TaeHun Kim}
\email[]{gimthcha@kias.re.kr}
\affiliation{School of Physics, Korea Institute for Advanced Study, 85 Hoegi-ro, Dongdaemun-gu, Seoul 02455, Republic of Korea}
\author{Philip Lu}
\email[]{philiplu11@gmail.com}
\affiliation{School of Physics, Korea Institute for Advanced Study, 85 Hoegi-ro, Dongdaemun-gu, Seoul 02455, Republic of Korea}


\begin{abstract}
Primordial black holes (PBH) can arise in a wide range of scenarios, from inflation to first-order phase transitions. Light PBHs, such as those produced during preheating, in bounce cosmologies, or at the GUT scale, could induce an early matter-dominated phase given a moderate initial abundance. During the early matter domination, the growth of initial PBH density perturbations can trigger collapse on horizon scales, producing much heavier PBHs. While the remaining original PBHs evaporate and reheat the Universe, these massive reformed PBHs survive for an extended period of time, producing potentially observable signatures at the present. We study this PBH reformation scenario and show that those reformed PBHs can emit significant quantities of gamma rays detectable by the next generation of experiments. The rapid reheating after matter domination generates a coincident stochastic gravitational wave background, which could be within the range of the upcoming CMB-S4 experiment. The PBH reformation scenario provides an intriguing possibility of decoupling the current PBH population and the initial formation mechanism from early Universe physics, while providing opportunities for observation through multi-messenger astronomy. 
\end{abstract}

\maketitle

\allowdisplaybreaks


\section{Introduction}

Primordial black holes (PBHs) can be produced in a variety of new physics scenarios, ranging from the collapse of perturbations seeded by inflationary fluctuations~\cite{Carr:1974nx,Carr:1975qj,Garcia-Bellido:1996mdl,Yokoyama:1995ex,Garcia-Bellido:2017mdw,Musco:2020jjb} to first-order phase transitions~\cite{Hawking:1982ga,Moss:1994iq,Khlopov:1998nm,Jung:2021mku,Hong:2020est,Kawana:2021tde,Lu:2022paj,Kawana:2022lba,Lu:2022jnp,Lu:2022yuc,Marfatia:2024cac,Liu:2021svg,Kawana:2022olo} and many others~\cite{Hawking:1987bn,Cotner:2018vug,Conzinu:2020cke,Helfer:2016ljl,Ruffini:1969qy}. The conceivable mass range of PBHs span more than 40 orders of magnitude, from the Planck mass to supermassive black holes. Of particular interest are light PBHs with masses between $5\times10^{14} \textrm{g}<M_{\rm PBH}\lesssim10^{17} \textrm{g}$, which can emit copious amounts of high energy Hawking radiation~\cite{Hawking:1974rv,Hawking:1976de,Page:1976df}. Consequently, this parameter space has been constrained by a variety of experiments~\cite{Carr:2009jm,Carr:2016hva,Boudaud:2018hqb,Huang:2024xap,Kim:2020ngi,Laha:2020vhg} searching for this emission. The next generation of observatories~\cite{LHAASO:2019qtb,CTAConsortium:2017dvg,Albert:2019afb} will further improve the sensitivity to high energy $\gamma$ and cosmic rays, enabling them detect a small but significant population of evaporating PBHs.

PBHs with masses below $5\times10^{14} \textrm{g}$ are usually considered to have completely evaporated by the present day~\cite{Page:1976df,MacGibbon:1990zk} and could be detectable by their traces on the Cosmic Microwave Background (CMB) and nuclide abundances from Big Bang Nucleosynthesis (BBN)~\cite{Carr:2009jm,Acharya:2020jbv,Chluba:2020oip}. These bounds extend down to $M\gtrsim 10^{9} \textrm{g}$, below which the PBH lifetime is short enough to evaporate before the epoch of BBN $\sim 1 \textrm{s}$. Recently, a series of papers~\cite{Inomata:2020lmk,Papanikolaou:2020qtd,Domenech:2020ssp,Domenech:2021wkk,Papanikolaou:2022chm,Domenech:2024wao} have shown that extremely light PBHs could usher in an era of early matter-domination (eMD) and source a large gravitational wave (GW) background when they evaporate, during the rapid transition from matter-domination to radiation-domination (RD). Although generally undetectable by present-day GW experiments, these GWs would constitute an excess contribution to the radiation energy density. As a result, $\Delta N_{\rm eff}$ limits derived from CMB and BBN observations constrain the abundance of PBHs down to $M\gtrsim10^{-2} \textrm{g}$~\cite{Domenech:2020ssp,Domenech:2021wkk}.

In this paper, we consider PBH reformation during and after the eMD era. Shortly after matter-radiation equality, the overdensities associated with the PBHs grow and could become substantial before they evaporate. During the eMD phase, the probability of PBH reformation is enhanced, following a power law~\cite{Khlopov:1980mg,1981SvA....25..406P,Harada:2016mhb,Harada:2017fjm,Kokubu:2018fxy} instead of the usual exponential dependence~\cite{Carr:1975qj,Young:2014ana}. This enables the possibility that collections of PBHs may collapse and form much larger PBHs. A similar concept was explored in Ref.~\cite{DeLuca:2022bjs}, where they discussed mergers of solar mass PBHs into supermassive black hole seeds, enhanced by strong clustering. In contrast, we focus on very light PBHs that would otherwise evaporate, and could initiate an eMD era in the early Universe, with collapse of a PBH gas rather than individual mergers as the formation mechanism. These reformed PBHs would be much lighter, actively emitting Hawking radiation which could be detected along with the coincident GW signal. 
This reformation framework is similar to PBH formation from other compact objects, such as Q-balls~\cite{Cotner:2016cvr, Cotner:2017tir} and oscillons~\cite{Cotner:2018vug, Cotner:2019ykd}, with sufficient initial density to initiate an eMD phase and decay rate for the initial population to evaporate.

We start by reviewing the relevant quantities and scales for PBH formation and the onset of an eMD era in Sec.~\ref{sec:PBHparam}. The power spectrum and its growth is discussed in Sec.~\ref{sec:powerspectrum} which leads to the PBH reformation during eMD in Sec.~\ref{sec:matterreform}. The accompanying GW background is discussed in Sec.~\ref{sec:GW}. We then discuss the additional possibility of reformation during the RD phase after PBH evaporation in Sec.~\ref{sec:evapreform}. Finally, we conclude in Sec.~\ref{sec:conclusions}.

\section{PBH Parameters}
\label{sec:PBHparam}

The mass of PBHs formed in many scenarios, such as from overdensities seeded by inflation and some involving phase transitions, is related to the horizon mass at formation by
\begin{equation}
\label{eq:masstemprelation}
    M_{\rm PBH} = \gamma\frac{4\pi M_P^2}{H} \simeq  47~\textrm{g}~\left(\frac{\gamma}{0.5}\right) \left(\frac{T}{10^{15}\textrm{ GeV}}\right)^{-2}~,
\end{equation}
where we take $\gamma=0.5$ as a fiducial value~\cite{Carr:2020xqk} for both the initial formation and the reformation. Unless otherwise specified, this relation is used for the following calculations, but should be modified if the initial formation scenario involves subhorizon scales (e.g. from bubble collisions~\cite{Hawking:1982ga,Moss:1994iq,Jung:2021mku} or compact remnants~\cite{Hong:2020est,Kawana:2021tde,Kawana:2022lba,Lu:2022paj}).

At their formation, the fraction of PBHs with respect to the overall energy density is characterized by $\beta=\rho_{\rm PBH}/\rho_{\rm tot}$, where $\rho_{\rm tot}$ is the total energy density of the PBHs and the relativistic SM plasma. Here we differentiate between the density fraction of PBHs at initial formation, $\beta_{\rm if}$, and the density fraction at reformation $\beta_{\rm rf}$. We focus on initial formation masses, $M_{\rm pl}<M_{\rm if} \lesssim 4\times10^{8} \textrm{g}$, which are unconstrained by BBN and CMB considerations. Such light PBHs are typical of GUT-scale formation~\cite{Anantua:2008am,Zagorac:2019ekv}, bounce cosmology~\cite{Papanikolaou:2023crz,Papanikolaou:2024fzf}, and preheating scenarios~\cite{Bassett:2000ha,Green:2000he,Martin:2019nuw,Ballesteros:2024hhq}.

Since PBHs are non-relativistic objects, their density increases relative to the SM plasma with redshift, which can result in an eMD phase. The matter(PBH)-radiation equality scale factor is simply related to the scale factor at initial formation by $a_{\rm eq} = a_{\rm if}/\beta_{\rm if}$. Likewise, the Hubble rate at the equality is proportional to the Hubble rate at initial formation, and the comoving horizon wavenumber is given by
\begin{equation}
\label{eq:keq}
    k_{\rm eq} = a_{\rm eq} H_{\rm eq} = \sqrt{2}a_{\rm if} H_{\rm if} \beta_{\rm if}~.
\end{equation}
Here, the subscript ``eq'' refers to the PBH-radiation equality as opposed to the matter-radiation equality around the epoch of the CMB. Light PBHs emit significant Hawking radiation, evaporating in a short timescale,
\begin{equation}
    \tau_{\rm evap} = 4.0\times10^{-4} \textrm{s} \left(\frac{M_{\rm PBH}}{10^8 \textrm{ g}}\right)^3\left(\frac{108}{g_H}\right)~,
\end{equation}
where the Hawking radiation degrees of freedom $g_H\simeq 108$ incorporates all the SM modes for $M_{\rm PBH} \lesssim 10^{13} \textrm{g}$. 

For eMD to occur, the initial density fraction must be large enough so that PBHs can dominate the energy density before they evaporate,
\begin{equation}
\label{eq:eMDcondition}
    \beta_{\rm if} > 4\times10^{-6} \left(\frac{M_{\rm PBH}}{1 \textrm{ g}}\right)^{-1} \left(\frac{g_*(T_{\rm RH})}{g_*(T_{\rm if})}\right)^{1/3}~,
\end{equation}
where $g_*$ is the entropic relativistic degrees of freedom in the SM plasma. The evaporation of the PBHs ends the eMD, reheating the universe to a temperature
\begin{eqnarray}
\label{eq:Trh}
    T_{\rm RH} &\simeq& 2.8\times10^{10}\textrm{ GeV} \nonumber \\
    &&\times \left(\frac{M_{\rm PBH}}{1~\textrm{g}}\right)^{-3/2} \left(\frac{g_*(T_{\rm RH})}{106.75}\right)^{-1/4}\left(\frac{g_H}{108}\right).
\end{eqnarray}
The comoving wavenumber at this reheating scale is
\begin{eqnarray}
    k_{\rm RH} &=& a_{\rm RH} H_{\rm RH} \nonumber \\
    &=& 2\times10^{-4} \, a_{\rm if} H_{\rm if}\left(\frac{\beta}{\gamma}\right)^{1/3}  \left(\frac{M_{\rm PBH}}{1~\textrm{g}}\right)^{-2/3}.
\end{eqnarray}
If the universe is matter-dominated for a significant period of time before evaporation, the tensor modes from inflationary reheating would be suppressed by the redshift during this period.

\section{Power Spectrum}
\label{sec:powerspectrum}

The mean separation between PBHs at the initial formation, $\Bar{r}$, defines the comoving ultraviolet cut-off wavenumber~\cite{Papanikolaou:2020qtd,Domenech:2020ssp},
\begin{equation}
    k_{\rm UV} = \frac{a}{\Bar{r}} = \left(\frac{\beta_{\rm if}}{\gamma}\right)^{1/3}a_{\rm if}H_{\rm if}~. \label{eq:kUV}
\end{equation}
For $k<k_{\rm UV}$, the PBHs can be treated as a Poissonian fluid with initial density power spectrum~\cite{Papanikolaou:2020qtd}
\begin{equation}
    \mathcal{P}_{\delta}(t_{\rm if})= \frac{2}{3\pi}\left(\frac{k}{k_{\rm UV}}\right)^3~. \label{eq:powerspectruminitial}
\end{equation}
During the eMD, PBH density perturbations grow linearly with scale factor~\cite{1974A&A....37..225M} once they cross the horizon. On the other hand, the growth of short wavelength modes that enter during RD is delayed until matter-radiation equality. 

The PBH density perturbation starts as an isocurvature mode at the moment of initial formation. But it becomes the adiabatic perturbation once PBHs dominate the Universe, similar to the curvaton scenario~\cite{Linde:1996gt, Enqvist:2001zp, Lyth:2001nq, Moroi:2001ct}. Therefore, PBH reformation from the blue-tilted power spectrum in Eq.~\eqref{eq:powerspectruminitial} is comparable to PBH formation in curvaton models with substantial small-scale power~\cite{Ando:2017veq,Gomez:2020rqv}.

The evolution of the power spectrum is described by the transfer function, $\mathcal{P}_\delta(t) = \mathcal{T}^2(t) \mathcal{P}_\delta(t_{\rm if})$. We numerically computed the linear transfer function under the relevant initial conditions (see Supplemental Material), and found a semi-analytic fit deep in the eMD phase,
\begin{align}
\label{eq:Transferanalyticapprox}
    \mathcal{T}(t) &\simeq \left( \frac{0.267 \, \kappa^2 + 1.5 \, \kappa^4}{\kappa^4 + 7.25 \kappa^3 + 4 \kappa^2 + 1.25 \kappa + 1} \right) y \quad \\ 
    &\simeq
        \begin{cases}
    	\frac{3}{2} \, y, & \text{for $\kappa \gg 1$}\\
            \frac{4}{15} \, \kappa^2 y, & \text{for $\kappa \ll 1$} 
        \end{cases}~.\nonumber
\end{align}
Here $\kappa = k / k_{\rm eq}$ and $y = a / a_{\rm eq}$ are the wavenumber and scale factor with respect to the matter-radiation equality, and $y \gg 1$ is assumed. 

The variance $\sigma(r,t)$, which measures the size of fluctuations at a given scale $r$ and time $t$, can be found by integrating the power spectrum,
\begin{equation}
    \sigma^2_{\delta} (r,t) = \int^\infty_0 \mathcal{P}_{\delta}(t) W^2(kr) \frac{dk}{k} \label{eq:sigmasqdelta}~.
\end{equation}
We use the real space top-hat window function, which captures the fluctuations within a spherical volume of radius $r$ and suppresses contributions from long wavelength modes with $k\ll r^{-1}$~,
\begin{equation}
    W(kr) = 3 \frac{\sin (kr) - kr \cos (kr)}{(kr)^3}. \label{eq:windowfunc}
\end{equation}
The dominant contribution to the variance in Eq.~\eqref{eq:sigmasqdelta} comes from the modes around $k_{\rm eq}$. During the matter domination phase or after reheating, this lies well inside the horizon where the window function has asymptotic behavior
\begin{equation}
    W^2(kr \gg 1) \simeq \frac{9}{2} \frac{1}{(kr)^4} \simeq \frac{9}{8} \frac{1}{\kappa^4 y^2}~ \label{eq:windowfuncapprox}
\end{equation}
by averaging out the oscillation. Substituting these approximations, Eqs.~\eqref{eq:Transferanalyticapprox} and~\eqref{eq:windowfuncapprox}, into the variance integral at the reheating scale, the integrand scales as
\begin{equation}
    P_\delta(t) W^2(kr) \propto
    \begin{cases}
		\kappa^{-1}, & \text{for $\kappa > 1$}\\
        \kappa^3, & \text{for $\kappa < 1$}
    \end{cases} \label{eq:varianceapprox}~
\end{equation}
with time dependence canceled for a long enough period of eMD.

\begin{figure}
    \centering
    \includegraphics[width = 1\linewidth]{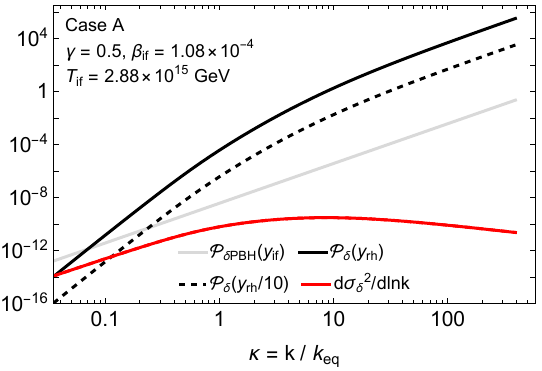}
    \caption{The initial PBH power spectrum (gray), power spectrum at the moment of reheating (black) and at $y = y_{\rm rh}/10$ (black dashed), and the variance at the horizon scale per logarithmic interval of wavenumber (red). The initial power spectrum is given by the initial Poisson distribution of PBHs, which then grows according to the transfer function, and then multiplied by the window function to give the variance. The parameters are chosen to produce a sizeable population of reformed PBHs, represented by Case A in Table~\ref{tab:mdformation} and Fig.~\ref{fig:reformedPBH}. All four lines are truncated at the UV cutoff $k_{\rm UV}$ in Eq.~\eqref{eq:kUV}.}
    \label{fig:powerspectrum}
\end{figure}

In Fig.~\ref{fig:powerspectrum}, we show the power spectrum at the initial formation time and during the eMD era based on linear analysis. As in Eq.~\eqref{eq:varianceapprox}, the variance per logarithmic wavenumber intervals is constant in time and levels off at the matter-radiation equality scale $k\simeq k_{\rm eq}$. If any initial clustering of PBHs exists, the power at small scales increases and the contribution to the variance continues to increase for larger wavenumbers, making the behavior around the ultraviolet cutoff dominate the reformation. 

At large $k$ deep in the matter-dominated era, the linear power spectrum at subhorizon scales can reach $P_\delta \gtrsim \mathcal{O}(1)$ which marks the breakdown of the linear theory. Just like particle dark matter, PBHs will typically form halos at those scales (see Ref.~\cite{Holst:2024ubt} for another reformation scenario through mergers in these halos), undergoing non-linear collapse and increasing in density by an $\mathcal{O}(100)$ factor~\cite{Desjacques:2016bnm}. We neglect these non-linear effects, which may increase the overdensity at horizon scales, and instead use linear theory to obtain a conservative estimate of the PBH reformation probability. Although there are well-known methods~\cite{Peebles:1980yev, Hamilton:1991es,Mo:1995db,Peacock:1996ci,Smith:2002dz,Takahashi:2012em,Bartlett:2024jes} to analytically transform the linear power spectrum into the non-linear power spectrum, we find that they do not yield reasonable results in this application. We leave non-linear studies to future works.

\section{PBH Reformation in \MakeLowercase{e}MD}
\label{sec:matterreform}

Refs.~\cite{Khlopov:1980mg,1981SvA....25..406P,Harada:2016mhb,Harada:2017fjm,Kokubu:2018fxy} considered the possibility of PBH formation during MD. In the absence of pressure forces, the critical overdensity threshold may be expected to tend toward zero~\cite{Musco:2012au}. However, two additional considerations of isotropy and homogeneity are required to ensure gravitational collapse. Instead of forming a PBH, an anisotropic perturbation undergoes a ``pancake'' collapse, forming caustics and resulting in a virialized configuration. Using the anisotropy distribution of Ref.~\cite{1970Ap......6..320D}, the fraction of perturbations that are sufficiently spherically symmetric enough to collapse into a PBH was computed by Ref.~\cite{Khlopov:1980mg,1981SvA....25..406P,Harada:2016mhb}, well approximated by $\beta = 0.05556\sigma_\delta^5$. These exceptional horizon-scale perturbations avoid virialization and halo formation.

Additionally, these perturbations must be homogeneous enough to form a horizon (i.e. satisfy the hoop criterion) before forming a ``naked singularity" which signifies the breakdown of the calculation. This limitation on the structure is satisfied by only a fraction of perturbations $\beta = 3.70\sigma_\delta^{3/2}$~\cite{Kokubu:2018fxy}, but may be model dependent~\cite{Harada:2016mhb}. Taken independently, these two conditions would suggest a minimum PBH formation probability proportional to $\beta \propto \sigma^{13/2}$. However, both conditions are more easily satisfied by larger perturbations, which can shrink to their larger Schwarzchild radii before becoming too anisotropic or inhomogenous. Since these conditions are correlated and dependent, the actual conditional probability should then be much higher than the minimum estimate. Therefore, we use the optimistic estimate of $\beta \simeq 0.05556 \sigma^5$ in this paper, which assumes that perturbations satisfying the much more stringent constraint of isotropy are also homogeneous enough. We ignore spin and angular momentum effects~\cite{Harada:2017fjm} which may alter the probability of PBH formation for small values of $\sigma_\delta$.

Significantly, the density of PBHs formed during MD follows a power law in $\sigma_\delta$ as opposed to the erfc dependence for RD, resulting in exponentially greater production for $\sigma_\delta \lesssim 10^{-2}$. This allows a modest variance $\sigma_\delta \sim 10^{-4}-10^{-3}$ during eMD to produce populations of reformed PBHs that contribute appreciably to the dark matter density, Eq.~\eqref{eq:fpbhbeta}, resulting in observable signatures.

\begin{table}[t]  
\centering 
\begin{tabular}{|c|c|c|c|c|}  
\hline 
Case & $T_{\rm if}$ (GeV) & $\beta_{\rm if}$ & $\gamma$ & $f_{\rm PBH}$ \\
\hline
A & $2.88\times10^{15}$ & $1.08\times 10^{-4}$ & 0.5 & $2.40\times10^{-5}$\\
\hline
B & $5.89\times10^{14}$ & $1.45\times 10^{-5}$ & 0.5 & $9.05\times10^{-12}$\\
\hline
\end{tabular}
\caption{
Example cases where PBH reformation during eMD results in a PBH population detectable through upcoming experiments. Reformation at successive scales during eMD results in flat PBH populations shown in Fig.~\ref{fig:reformedPBH}.
}
\label{tab:mdformation}
\end{table}

\begin{figure}[t]
    \centering  
    \includegraphics[width = 1\linewidth]{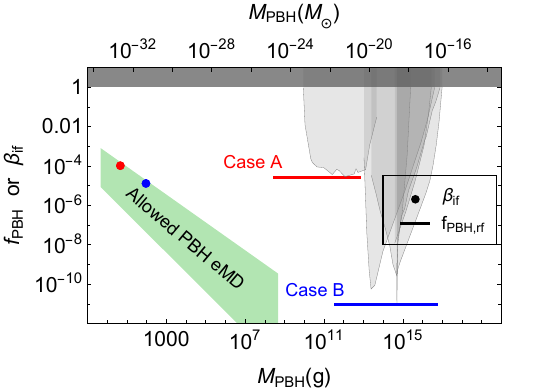}
    \caption{Two populations of PBHs reformed during the eMD era that could be detected in the next generation of experiments. Both Case A (red) and Case B (blue) are in the allowed region of initial PBH parameters (green region) where the initial density $\beta_{\rm if}$ is large enough to result in an eMD phase and below the GW bound. The PBH spectrum for Case A lies below the current bounds from BBN and the emission of PBHs reformed in Case B could be detected by the next generation of CMB and $\gamma$-ray experiments. The gray-shaded regions are constraints from BBN~\cite{Carr:2009jm}, CMB~\cite{Acharya:2020jbv, Chluba:2020oip}, extragalactic $\gamma$-rays~\cite{Carr:2009jm}, galactic $\gamma$-rays~\cite{Carr:2016hva}, Voyager-1 $e^\pm$~\cite{Boudaud:2018hqb}, and AMS-02~\cite{Huang:2024xap}.
    }
    \label{fig:reformedPBH}
\end{figure}

We apply the formula of Ref.~\cite{Harada:2016mhb} to PBH reformation during the eMD, from matter-radiation equality up until evaporation and reheating. Evaluating the variance at the horizon size of each epoch by Eqs.~\eqref{eq:sigmasqdelta} and \eqref{eq:varianceapprox}, we find it to be roughly constant and approximated by
\begin{equation}
    \sigma_\delta^2 \simeq \frac{0.06}{\kappa_{\rm UV}^3} - \frac{0.54}{\kappa_{\rm UV}^4} + \frac{3.9}{\kappa_{\rm UV}^5} - \frac{27}{\kappa_{\rm UV}^6} + \cdots
\end{equation}
for $\kappa_{\rm UV} = 1/\sqrt{2}\gamma^{1/3}\beta_{\rm if}^{2/3} \gg 1$. Here, $\kappa_{\rm UV}$ dependence comes from the fixed value of $\mathcal{P}_\delta (t_{\rm if}) (k_{\rm UV}) = 2/3\pi$ (Eq.~\eqref{eq:powerspectruminitial}) which makes $\mathcal{P}_\delta (t_{\rm if}) (k_{\rm eq})$ to be suppressed by the separation between $k_{\rm eq}$ and $k_{\rm UV}$, as in Fig.~\ref{fig:powerspectrum}.
Unlike in the radiation-dominated era, where the PBH density grows relative to the radiation plasma immediately after formation, these reformed PBHs maintain the same energy density fraction during eMD. Thus a constant $\beta_{\rm rf}$ translates to a constant $f_{\rm PBH}$, with the relation
\begin{equation}
\label{eq:fpbhbeta}
    f_{\rm PBH} = 4.0\times10^{20} \beta \left(\frac{M_{\rm if}}{1~\textrm{g}}\right)^{-3/2}\left(\frac{g_*(T_{\rm RH})}{106.75}\right)^{1/12}~.
\end{equation}
This flat mass spectrum has endpoints defined by the horizon scales at matter-radiation equality and reheating, $k_{\rm RH} < k < k_{\rm eq}$. From Eqs.~\eqref{eq:masstemprelation},~\eqref{eq:keq}, and~\eqref{eq:Trh}, the mass range of the reformed PBH spectrum is 
\begin{equation}
    \frac{M_{\rm if}}{\sqrt{2} \beta_{\rm if}^2} \leq M_{\rm RH} \leq 6.0\times10^{10} \, \textrm{g} \left(\frac{\gamma}{0.5} \right) \left(\frac{M_{\rm if}}{1~\textrm{g}}\right)^3~.
\end{equation}
We list two example cases of PBH reformation in Table~\ref{tab:mdformation}, which generate interesting PBH populations below current experimental bounds and potentially detectable by upcoming observations.
In Fig.~\ref{fig:reformedPBH}, we display these two populations of PBHs reformed during the eMD era using the parameters in Table~\ref{tab:mdformation}. The initial PBH density, $\beta_{\rm if}$, in both sets are sufficient to produce an eMD phase, and are also extremely light so that they evaporate well before BBN.

For Case A, the reformed PBH population are at the cusp of current BBN constraints from observations of the primordial deuterium abundance~\cite{osti_5338600,10.1093/mnras/193.3.593,Ellis:1990nb}. Experiments targeting the processes relevant for deuterium production could further reduce prediction uncertainties and lead to either more stringent constraints or an anomaly~\cite{Mossa:2020gjc,Xu:2023jyr,Shen:2024}. The reformed PBH population of Case B have lifetimes comparable to that of the Universe $\sim 10^{10}$ yr, and could be emitting high energy particles ($\gamma$,$e^{\pm}$) at the present. The evaporation signal of these reformed PBH may be within the sensitivity limit of the operational run of LHAASO~\cite{LHAASO:2019qtb,Yang:2024vij} and next generation $\gamma$-ray experiments such as CTA~\cite{CTAConsortium:2017dvg} and SGSO~\cite{Albert:2019afb}. We consider a possible coincident GW signature in the next section.

\section{Gravitational Wave Signal}
\label{sec:GW}

The rapid evaporation and reheating after the eMD phase leads to a coincident GW signal. The stochastic GW background emitted during the PBH reheating was computed in Refs.~\cite{Inomata:2020lmk,Papanikolaou:2020qtd,Domenech:2020ssp,Domenech:2021wkk,Papanikolaou:2022chm,Domenech:2024wao}, and showed that the rapid reheating characteristic of Hawking evaporation led to significant tensor mode production. The resulting GW spectrum is highly wavenumber dependent, $\Omega_{\rm GW}\sim k^{11/3}$ and peaks at the ultraviolet cutoff $k_{\rm UV}$~\cite{Domenech:2020ssp},
\begin{align}
\begin{split}
    \Omega_{\rm GW,0}(k_{\rm UV}) \simeq& \, 2.1\times10^{-10}\left(\frac{k_{\rm UV}}{k_{\rm RH}}\right)^{17/3}\left(\frac{k_{\rm eq}}{k_{\rm UV}}\right)^8 \\ 
    \simeq& \, 3.2\times10^{11}\beta_{\rm if}^{16/3}\left(\frac{\gamma}{0.5}\right)^{8/3}\left(\frac{M_{\rm PBH}}{1 \, \text{g}}\right)^{34/9}~.
\end{split}
\end{align}

Depending on the parameters, the stochastic GW background generated by PBH reheating may be detected by planned upgrades to aLIGO~\cite{abbott2020prospects} as well as upcoming experiments such as LISA~\cite{LISA:2017pwj}, Einstein Telescope~\cite{hild2011sensitivity}, and Cosmic Explorer~\cite{srivastava2022science}. Additionally, the emission of GWs can be constrained by the current bound on $\Delta N_{\rm eff} \lesssim 0.5$ by Planck TT+low E observations~\cite{Planck:2018vyg} and similarly by BBN observations~\cite{Arbey:2021ysg}, and be potentially detected by the CMB-S4 experiment which has a sensitivity limit of $\Delta N_{\rm eff} = 0.06$~\cite{Abazajian:2019eic}.

\begin{figure}
    \centering  
    \includegraphics[width = 1\linewidth]{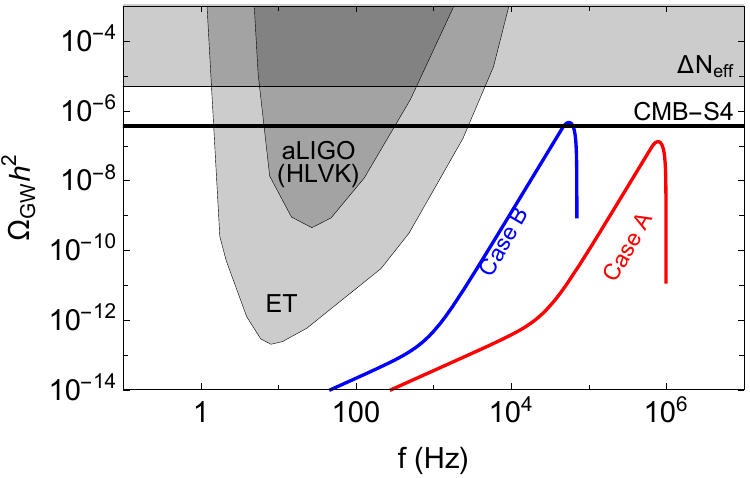}
    \caption{The coincident GW signature of two models of reformed PBH populations. The initial populations of extremely light PBHs result in a high frequency stochastic GW background after evaporation, higher than aLIGO or ET's reachable range~\cite{Schmitz:2020syl}. These GWs contribute to $\Delta N_{\rm eff}$ at BBN~\cite{Caprini:2018mtu} and could be detected by the upcoming CMB-S4 experiment~\cite{Abazajian:2019eic}. 
    }
    \label{fig:GWsignature}
\end{figure}

In Fig.~\ref{fig:GWsignature}, we show the resulting GW signatures for the same two cases A and B. These stochastic backgrounds lie just below the $\Delta N_{\rm eff}$ limits set by current CMB and BBN observations, and could potentially be detected by CMB-S4. The possible discovery of gamma rays from an evaporating PBH population correlated with a high frequency GW background would be suggestive of the PBH reformation process.

\section{PBH Reformation after Reheating}
\label{sec:evapreform}

After reheating, the inhomogenous PBH distribution is converted into radiation inhomogeneities. Horizon-sized overdense regions could then grow and recollapse into a PBH~\cite{Carr:1975qj}. We estimate the abundance of these reformed PBHs during RD using the formula~\cite{Young:2014ana,Carr:2020xqk}
\begin{equation}
\label{eq:betarad}
    \beta_{\rm rf} = \gamma~ \textrm{erfc}\left(\frac{\delta_c}{\sqrt{2}\sigma_\delta}\right)~,
\end{equation}
where the fiducial value for the critical collapse threshold is taken to be $\delta_c=0.45$~\cite{Musco:2012au,Carr:2020xqk} and the variance is evaluated at reheating $\sigma_\delta(k_{\rm RH}^{-1},t_{\rm RH})$. Due to the $\textrm{erfc}$ in Eq.~\eqref{eq:betarad}, the abundance of reformed PBHs depends sensitively on the ratio $\delta_c/\sigma_\delta$, with significant production occurring only for $\sigma \gtrsim \mathcal{O}(10^{-2})$.

This makes any significant reformation after reheating practically rely on the clustering of the initial PBHs. Here we remain agnostic about the clustering mechanism, and adopt a phenomenological parameterization of Ref.~\cite{Domenech:2020ssp}, which is convenient for both the computation of the GW signal and PBH reformation,
\begin{equation}
    P_{\delta,\rm cl}(k) = P_\delta(k)\left(\frac{k}{k_{\rm UV}}\right)^3 \left(1+F_{\rm cl}\left(\frac{k}{k_{\rm cl}}\right)^{n_{\rm cl}}\right)~,
\label{eq:clustering}
\end{equation}
where $F_{\rm cl}$ is the amplitude of clustering, $k_{\rm cl}$ is the clustering scale, and $0<n_{\rm cl}\leq 1$ is the range of power laws that we consider. The clustering enhances the power spectrum for $k > k_{\rm cl}$, and hence the reformation of PBHs after reheating. Correspondingly, the GW emission which is dominated by the UV cutoff also scales with the increased power at the UV scale, 
\begin{equation}
    \Omega_{\rm GW, cl, 0}(k_{\rm UV}) = F_{\rm cl}^2 \left(\frac{k}{k_{\rm cl}}\right)^{2n_{\rm cl}} \Omega_{\rm GW, 0}(k_{\rm UV})~.
\end{equation}
If these clustered PBHs induce the last eMD, right before the standard cosmological history,  the current $N_{\rm eff}$ bound results in a constraint on the $M_{\rm if},\beta$, and clustering parameters~\cite{Domenech:2020ssp,Domenech:2021wkk},
\begin{eqnarray}
    \beta_{\rm if} &\lesssim& 5.6\times10^{-4} F_{\rm cl}^{-3/8} \nonumber \\
    &&\times \left(\frac{k_{\rm cl}}{k_{\rm UV}}\right)^{3n_{\rm cl}/8}\left(\frac{\gamma}{0.5}\right)^{-1/2}\left(\frac{M_{\rm if}}{1~\textrm{g}}\right)^{-17/24}~. \quad
\end{eqnarray}

\begin{figure}
    \centering  
    \includegraphics[width = 1\linewidth]{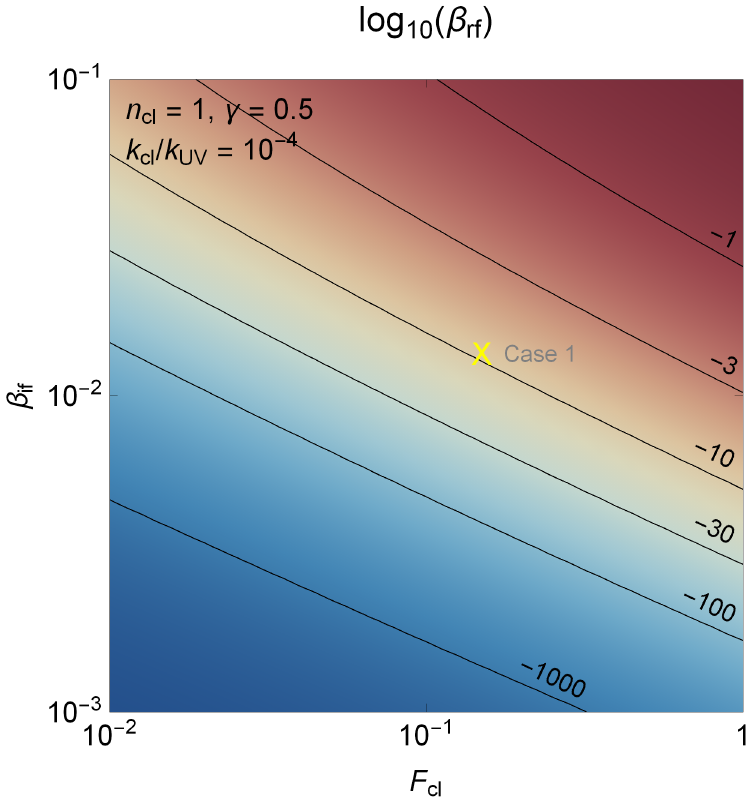}
    \caption{Contour plot showing the resulting PBH abundance $\beta_{\rm rf}$ varying the initial PBH density $\beta_{\rm if}$ and the clustering strength $F_{\rm cl}$, with other relevant parameters shown. Significant reformation of PBHs can occur after reheating for $\beta_{\rm if} \gtrsim 10^{-2}$ depending on the initial clustering parameters. A second round of eMD could occur, resulting in cascading PBH formation, like the yellow cross representing Case 1 in Table~\ref{tab:cascadescenario}.}
    \label{fig:contourplot}
\end{figure}

In Fig.~\ref{fig:contourplot}, we show the abundance of reformed PBHs, $\beta_{\rm rf}$, as a function of initial abundance $\beta_{\rm if}$ and clustering parameters. For initial PBH densities down to $\beta_{\rm if}\sim 10^{-2}$, significant quantities of reformed PBHs can be produced. These may be sufficient to induce a second eMD era, or even result in cascading production of increasingly heavy reformed PBHs. To avoid interfering with BBN processes, only reformed PBHs with mass $M\lesssim 4\times10^8 \, \text{g}$ can undergo a second eMD.

\begin{table}[t]  
\centering 
\begin{tabular}{|c|c|c|c|c|c|c|}  
\hline 
Case & $T_{\rm if}$ (GeV) & $\beta_{\rm if}$ & $\gamma$ & $F_{\rm cl}$ & $k_{\rm cl}/k_{\rm UV}$ & $n_{\rm cl}$  \\
\hline
1 & $2\times10^{16}$ & 0.0136 & 0.5 & 0.15 & $10^{-4}$ & 1  \\
\hline
2 & $5\times10^{15}$ & 0.0214 & 0.0311 & 1.0 & $10^{-4}$ & 1  \\
\hline
\end{tabular}
\caption{
Example cases where the PBH reformation after reheating induces 2nd eMD. Both cases result in the reformed PBH population of mass $M_{\rm PBH,rf} \simeq 10^8 \, \text{g}$ and $\beta_{\rm rf} \simeq 10^{-9}$, being at the boundary of the allowed last PBH eMD (green shaded region of Fig.~\ref{fig:reformedPBH}). 
}
\label{tab:cascadescenario}
\end{table}

Table~\ref{tab:cascadescenario} lists the parameters for two example cases that result in reformed PBH populations of $M_{\rm PBH,rf} \simeq 10^8 \, \text{g}$ and $\beta_{\rm rf} \simeq 10^{-9}$. Case 1 assumes a typical horizon collapse during the radiation-dominated era right after the inflationary reheating, and Case 2 assumes a fast first-order phase transition at a very high temperature, producing subhorizon scale initial PBHs~\cite{Hawking:1982ga,Moss:1994iq,Jung:2021mku}. Although the formation temperature $T_{\rm if}$ of Case 1 slightly exceeds the bound on the inflation scale coming from the null detection of tensor modes in CMB observations, these excess GWs could be suppressed by redshift during the eMD phases. For Case 2, the PBHs do not strictly follow the mass-temperature relation, Eq.~\eqref{eq:masstemprelation}, which is possible in first-order phase transition formation scenarios.

\section{Conclusion}
\label{sec:conclusions}

A small population of extremely light PBHs with mass $M\lesssim 4\times10^8 \textrm{g}$ can engender an eMD era. The growth of structure during matter domination could seed the overdensities that collapse into larger, reformed PBHs. Our computation shows that PBH reformation is applicable to PBHs initially formed in a variety of scenarios, from inflationary fluctuations to first order phase transitions. We explored the possibility of PBH reformation during and right after the eMD era, which could result in evaporating reformed PBHs emitting high energy cosmic rays and gamma ray bursts. These signals could be matched with a coincident GW signature that results from the rapid evaporation and reheating of the initial PBH population. Correlated observations of these two phenomena would be indicative of the PBH reformation mechanism.

\acknowledgments
The authors thank Donghui Jeong for many helpful discussions and the referee for useful suggestions in improving the paper.
T.H.K. is supported by KIAS Individual Grant PG095201 at Korea Institute for Advanced Study.
P.L. is supported by KIAS Individual Grant 6G097701.

\appendix

\section{Numerical Computation of Linear Growth Equations} \label{app:Numerical}
We list the evolution equations for cosmological perturbations~\cite{Seljak:1996is} in the conformal Newtonian gauge. We have two scalar potentials, the hierarchical moments of radiation perturbation, the PBH density perturbation and velocity. In expressing the evolution, we use $\ln y = \ln (a / a_{\rm eq})$ as a time variable. The two scalar potentials evolve as
\begin{equation}
    \frac{d \Phi}{d \ln y} = \Psi - \frac{\kappa^2}{3} \frac{2y^2}{y+1}\Phi + \frac{1}{2} \left( \frac{y}{y+1} \delta_{\rm PBH} + \frac{1}{y+1} 4 \Theta_{r,0} \right)~, \label{eq:PhiEvol}
\end{equation}
with the relation
\begin{equation}
    \Psi = -\Phi - \frac{6}{\kappa^2 y^2} \Theta_{r,2}.
\end{equation}
The 0th moment of radiation perturbation follows
\begin{equation}
    \frac{d \Theta_{r,0}}{d \ln y} = - \frac{d\Phi}{d \ln y} - \kappa \sqrt{\frac{2y^2}{y+1}} \Theta_{r,1},
\end{equation}
and since the radiation scatters negligibly with the PBHs, higher moments follow
\begin{eqnarray}
    \frac{d \Theta_{r,1}}{d \ln y} &=& \frac{\kappa}{3} \sqrt{\frac{2y^2}{y+1}} (\Psi + \Theta_{r, 0} - 2 \Theta_{r, 2} )~, \\
    \frac{d \Theta_{r,2}}{d \ln y} &=& \frac{\kappa}{5} \sqrt{\frac{2y^2}{y+1}} (2 \Theta_{r,1} - 3\Theta_{r,2})~, \\
    \frac{d \Theta_{r,l}}{d \ln y} &=& \frac{\kappa}{2l+1} \sqrt{\frac{2y^2}{y+1}} [l \Theta_{r, (l-1)} - (l+1) \Theta_{r, (l+1)}] \qquad
\end{eqnarray}
with the Boltzmann hierarchy truncated at $l_{\rm max}$ through \cite{Ma:1995ey}
\begin{equation}
    \Theta_{r, (l_{\rm max} + 1)} \simeq \frac{2 l_{\rm max}}{2\sqrt{2} \kappa (\sqrt{1+y} - 1)} \Theta_{r, l_{\rm max}} - \, \Theta_{r, (l_{\rm max} - 1)}
\end{equation}
where we used $\eta = 2\sqrt{2} (\sqrt{1+y} - 1) / a_{\rm eq} H_{\rm eq}$ for the conformal time. The PBH density follows
\begin{equation}
    \frac{d \delta_{\rm PBH}}{d \ln y} = -3 \frac{d \Phi}{d \ln y} - i \kappa \sqrt{\frac{2y^2}{y+1}} \, v_{\rm PBH}
\end{equation}
and the velocity follows
\begin{equation}
    \frac{d v_{\rm PBH}}{d \ln y} = -i \kappa \sqrt{\frac{2y^2}{y+1}} \Psi - v_{\rm PBH}. \label{eq:vPBHevol}
\end{equation}

\begin{figure}[ht]
\includegraphics[width=0.6\linewidth]{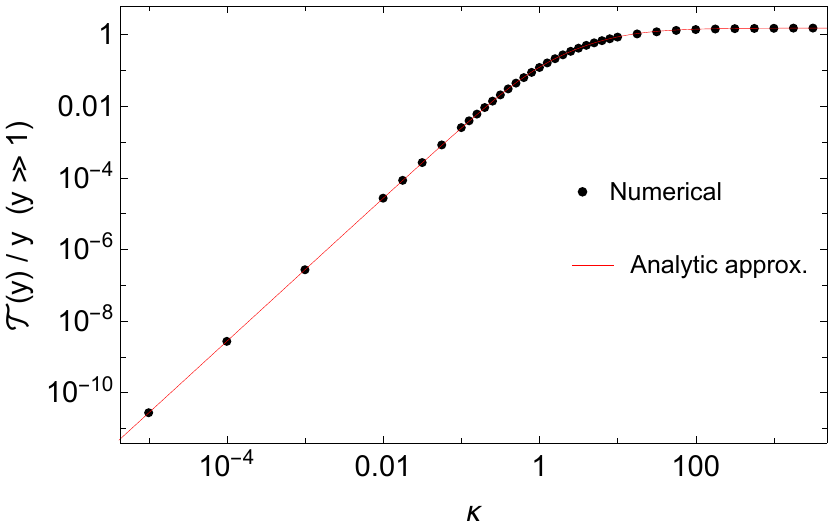}
\caption{Transfer function of density perturbation with numerical calculation (black dots) and analytic approximation in Eq.~(9) (red solid).
} 
\label{fig:Transferanalytic}
\end{figure}

To evaluate the transfer function, we numerically evolve Eqs.~\eqref{eq:PhiEvol}--\eqref{eq:vPBHevol} from $y\lesssim 10^{-5}$ to $y\gtrsim 10^{2}$ where the asymptotic behavior is reached. We implemented the initial condition with all vanishing initial perturbations except $\delta_{\rm PBH} = 1$, corresponding to inhomogeneously distributed PBHs on a homogeneous background. The Boltzmann hierarchy is truncated at $l_{\rm max} = 30$, but much smaller $l_{\rm max}$ gives nearly the same result. For $y \gg 1$, we have $\mathcal{T} \propto y$ and the result for selected $\kappa$'s are shown as black dots in Fig.~\ref{fig:Transferanalytic}. We heuristically found the fitting formula in Eq.~\eqref{eq:Transferanalyticapprox}, shown as the red solid curve in Fig.~\ref{fig:Transferanalytic}.

Since $\delta$ is a gauge-dependent variable, one must check whether any artifact comes from the gauge choice. However, the evolution in $y \gg 1$ shows that for subhorizon modes, only $\delta_{\rm PBH}$ grows as $\delta_{\rm PBH} \propto a$ while all other quantities vanish (modes entering the horizon before the equality) or remain a small $\mathcal{O}(0.1)$ value (modes entering the horizon after the equality). This practically renders the gauge problem for subhorizon modes unimportant. 

For superhorizon modes, we have constant values of $\delta_{\rm PBH} = 2/5$ and $\Psi = -1/5$ for $y \gg 1$. Since the two have comparable sizes, we use the gauge independent curvature perturbation $\zeta = -\Psi + \delta_{\rm PBH} / 3 = 1/3$. For superhorizon modes, the density perturbation that determines PBH formation is obtained by~\cite{Young:2019osy}
\begin{equation}
    \delta = \frac{2(1+w)}{5+3w} \left( \frac{k}{aH} \right)^2 \zeta
\end{equation}
with $w = 0$ for eMD and $w = 1/3$ for RD. Then, $\delta$ obtained in this way smoothly connects to $\delta$ for subhorizon modes and finally results in the approximation in Eq.~\eqref{eq:Transferanalyticapprox}.

\bibliography{references}

\end{document}